\documentclass[fleqn,10pt]{wlscirep}
\usepackage[utf8]{inputenc}
\usepackage[T1]{fontenc}
\usepackage{adjustbox}
\usepackage[nolist]{acronym}
\usepackage{dirtytalk}
\usepackage{xspace}
\makeatletter
\DeclareRobustCommand\onedot{\futurelet\@let@token\@onedot}
\def\@onedot{\ifx\@let@token.\else.\null\fi\xspace}

\makeatother
\usepackage{siunitx}  
\DeclareSIUnit{\sec}{sec}
\DeclareSIUnit\pixel{px}
\usepackage[capitalise, noabbrev]{cleveref}
\usepackage{float}
\newcommand{\crefSubFigRef}[2]{\crefformat{figure}{Figure~##2##1{#2}##3}%
\cref{#1}\crefformat{figure}{Figure~##2##1##3}}
\newcommand{\annos}[0]{12,424 }

\title{Pan-tumor CAnine cuTaneous Cancer Histology (CATCH) dataset}

\author[1,*]{Frauke Wilm}
\author[2]{Marco Fragoso}
\author[1]{Christian Marzahl}
\author[3]{Jingna Qiu}
\author[2]{Chloé Puget}
\author[2]{Laura Diehl}
\author[4]{Christof A. Bertram}
\author[2]{Robert Klopfleisch}
\author[1]{Andreas Maier}
\author[3,$\dag$]{Katharina Breininger}
\author[5,$\dag$]{Marc Aubreville}

\affil[1]{Pattern Recognition Lab, Department of Computer Science, Friedrich-Alexander-Universit\"at Erlangen-N\"urnberg, Erlangen, Germany}
\affil[2]{Institute of Veterinary Pathology, Freie Universit\"at Berlin, Berlin, Germany}
\affil[3]{Department Artificial Intelligence in Biomedical Engineering, Friedrich-Alexander-Universit\"at Erlangen-N\"urnberg, Erlangen, Germany
}
\affil[4]{Institute of Pathology, University of Veterinary Medicine, Vienna, Austria}
\affil[5]{Technische Hochschule Ingolstadt, Ingolstadt, Germany}

\affil[*]{corresponding author: Frauke Wilm (frauke.wilm@fau.de)}
\affil[$\dag$]{joint senior authors}

\begin{abstract}
Due to morphological similarities, the differentiation of histologic sections of cutaneous tumors into individual subtypes can be challenging. Recently, deep learning-based approaches have proven their potential for supporting pathologists in this regard. However, many of these supervised algorithms require a large amount of annotated data for robust development. We present a publicly available dataset of 350 whole slide images of seven different canine cutaneous tumors complemented by \annos polygon annotations for 13 histologic classes, including seven cutaneous tumor subtypes. In inter-rater experiments, we show a high consistency of the provided labels, especially for tumor annotations. We further validate the dataset by training a deep neural network for the task of tissue segmentation and tumor subtype classification. We achieve a class-averaged Jaccard coefficient of 0.7047, and 0.9044 for tumor in particular. For classification, we achieve a slide-level accuracy of 0.9857. Since canine cutaneous tumors possess various histologic homologies to human tumors the added value of this dataset is not limited to veterinary pathology but extends to more general fields of application.
\end{abstract}
\begin{document}

\flushbottom
\maketitle
\thispagestyle{empty}
\section*{Background \& Summary}
The skin and soft tissue are the most common anatomical sites for canine neoplasms~\cite{dobson2002} and the segmentation and classification of canine cutaneous tumors are routine tasks for veterinary pathologists. Especially different types of round cell tumors, which can have similar morphologies, are oftentimes hard to distinguish on standard histologic stainings~\cite{fernandez2005,bertram2018}. Tumor-specific \ac{ihc} stainings can support the pathologist in this regard but are considerably more expensive, time-consuming, and still might not provide reliable results for undifferentiated tumors~\cite{fernandez2005}. Deep learning-based algorithms can assist the pathologist in segmenting and classifying cutaneous tumors on standard \ac{he} staining and have successfully been applied in various works~\cite{salvi2021,thomas2021,jiang2020,campanella2019,halicek2019,hekler2019,arevalo2015}. These algorithms, however, are often criticized for requiring vast amounts of labeled training data~\cite{marcus2018}. Therefore, publicly available datasets have become increasingly popular, as they reduce annotation costs for recurring pathological research questions and improve the comparability of computer-aided systems developed on these datasets. 

Most existing open access datasets for segmentation in histopathology originated from computer vision challenges. \Cref{tab:datasets} provides a collection of recently published datasets. These datasets not only differ in the anatomical location of the tumor and thereby the annotation classes, but also in the labeling method used for annotating the image data. Datasets consisting of small image patches, with only one tissue class present, are usually labeled on image level, whereas datasets with complete \acp{wsi} are typically annotated with polygon contours. The CAMELYON~\cite{litjens2018}, the \acs{bach} (\acl{bach})~\cite{aresta2019}, and the \acs{bracs} (\acl{bracs})~\cite{pati2022} dataset addressed lesion detection and classification for breast cancer and provide a mixture of image-level and contour annotations. Whereas the CAMELYON challenge focused on the detection of metastatic regions as a binary task, the latter two were designed for the classification into normal tissue and multiple lesion subtypes. The \acs{paip} (\acl{paip})~\cite{kim2021} \ac{wsi} dataset addressed the detection of neoplasms in liver tissue as a binary segmentation task. In contrast to the aforementioned datasets, which focused on lesion detection and classification in a specific tumor region, the \acs{adp} (\acl{adp})~\cite{hosseini2019} and the \acs{droid} (\acl{droid})~\cite{stadler2021} include images from multiple organs. Furthermore, they significantly exceed comparable datasets in terms of annotation classes. Whereas the \acs{adp} provides small tissue-specific patches labeled on image level, the \acs{droid} provides extensive polygon annotations on \acp{wsi}.

\begin{table}[H]
\centering
\begin{adjustbox}
{ width=\textwidth}
\begin{tabular}{|lllS[table-format = 5.0]lS[table-format = 2.0]ll|}
\hline
Dataset  & Year & Organ & \multicolumn{1}{l}{Images} & Image Type & \multicolumn{1}{l}{Classes} & Class Details &  \multicolumn{1}{l|}{Annotation Type}  \\
\hline
CAMELYON16 \cite{litjens2018} & 2016 & breast & 399 & \aclp{wsi} & 2 & tumor types & contours \\
CAMELYON17 \cite{litjens2018} & 2017 & breast & 1000 & \aclp{wsi}  & 5 & tumor stagings & image-level (all images)\\  
& & & & & 3 & tumor types &  contours (50 images) \\
\acs{bach} \cite{aresta2019} & 2018 & breast & 500 & patches & 4 & tumor types & image-level \\ 
 & & & 40 & \aclp{wsi} &  4 & tumor types & contours (10 images)\\ 
\acs{paip} \cite{kim2021} & 2019 & liver & 100 & \aclp{wsi} & 2 & whole tumor + viable tumor & contours \\
\acs{adp} \cite{hosseini2019} & 2019 & multi-organ & 17668 & patches & 57 & benign tissues & image-level \\ 
\acs{bracs} \cite{pati2022} & 2021 & breast & 4537 & patches & 7 & tumor types & image-level\\ 
& & breast & 574 & \aclp{wsi} & 7 & tumor types & image-level\\ 
\acs{droid} \cite{stadler2021} & 2021 & ovarian & 193 & \aclp{wsi} & 11 & ovarian tissue + tumor types & contours\\
& & breast & 361 & \aclp{wsi} & 4 & tumor types & contours\\
& & skin  & 99 & \aclp{wsi} & 32 &  benign + abnormal tissues & contours\\
& & colon  & 101 & \aclp{wsi} & 38 & benign + abnormal tissues & contours\\
\acs{catch} (ours) & 2022 & skin & 350 & \aclp{wsi} & 13 & skin tissue + tumor types & contours\\ 
\hline
\end{tabular}
\end{adjustbox}
\caption{\label{tab:datasets}Publicly available datasets for segmentation tasks on histological specimens.}
\end{table}

\noindent In this work, we present a dataset of \num{350} \acp{wsi} of seven canine cutaneous tumor subtypes, which we have named \ac{catch} dataset. As opposed to human samples, veterinary datasets are less affected by data-privacy concerns, which makes them more suited for public access. Furthermore, previous work has demonstrated homologies between canine and human cutaneous tumors~\cite{prouteau2019,ranieri2013,pinho2012} which supports the relevance of publicly available databases for both species. We provide contour annotations for six tissue classes and seven tumor subtypes. With \annos annotations and \num{13} classes, this dataset exceeds most publicly available datasets in annotation extent and label diversity. We validated annotation quality by evaluating the inter-observer variability of three pathologists on a subset of the presented dataset with high concordance for most annotation classes. Furthermore, we present results for two computer vision tasks on the presented dataset. We first segmented the \ac{wsi} into background, tumor, and the four most prominent tissue classes (epidermis, dermis, subcutis, and a joint class of inflammation and necrosis). We evaluated the segmentation result using the class-wise Jaccard coefficient, resulting in an average score of \num{0.7047} on our test set. Afterward, we classified the predicted tumor regions into one of seven tumor subtypes, achieving a slide-level accuracy of \SI{98.57}{\percent} on the test set. These results, achieved by standard architectures, are the first published results of computer vision algorithms trained on the \ac{catch} dataset and can serve as a baseline for the development of more complex architectures or training strategies. Furthermore, the successful training of these architectures validates dataset consistency. The dataset, as well as the annotation database, is publicly available on \ac{tcia}~\cite{catch}. Code examples for the methods presented in this work, along with a slide-level overview of the train-test split used for model development, can be obtained from our GitHub repository (\href{https://github.com/DeepPathology/CanineCutaneousTumors}{https://github.com/DeepPathology/CanineCutaneousTumors}).    

\section*{Methods}
\subsection*{Sample selection and preparation}
In total, 350 cutaneous tissue samples from 282 canine patients were selected retrospectively from the biopsy archive of the Institute for Veterinary Pathology of the Freie Universität Berlin. Use of these samples was approved by the local governmental authorities (State Office of Health and Social Affairs of Berlin, approval ID: StN 011/20). All specimens were submitted by veterinary clinics or surgeries for routine diagnostic examination of neoplastic disease. As to local regulations, no ethical vote is required for these samples. No additional harm or pain was induced in the course of this study. Samples were chosen uniformly from seven of the most common canine cutaneous tumors, according to pathology reports. The case selection was guided by sufficient tissue preservation and the presence of characteristic histologic features for the corresponding tumor subtypes. Samples from the same canine patient were obtained from spatially separated sections of the same tumor or different neoplasms of the same subtype. All samples were routinely fixed in formalin, embedded in paraffin, and tissue sections were stained with \ac{he}. \num{303} of the sections were digitized with the Leica ScanScope CS2 linear scanning system at a resolution of \SI{0.2533} (\num{40}X objective lens). Due to practical feasibility, \num{47} slides were digitized with a different, but very similar scanning system (Leica AT2) at the same magnification and a resolution of \SI{0.2524}{\micro\meter\per\pixel} (\num{40}X objective lens).

\subsection*{Annotation workflow}
All \acp{wsi} were annotated using the open source software SlideRunner~\cite{aubreville2018}. The \acp{wsi} were predominantly (\SI{82}{\percent}) annotated by the same pathologist (M.F.). The remaining annotations were gathered by three medical students in their 8th semester who were supervised by the leading pathologist (M.F.). M.F. later reviewed these annotations for correctness and completeness. Overall, annotations were gathered for seven canine cutaneous tumor subtypes as well as six additional tissue classes: epidermis, dermis, subcutis, bone, cartilage, and a joint class of inflammation and necrosis. The open source online platform EXACT~\cite{marzahl2021} was used to monitor slide and annotation completeness.

\section*{Data Records}
We provide public access to the full-resolution dataset on \ac{tcia}~\cite{catch}. In total, the dataset consists of \num{350} \acp{wsi} -- 50 each for seven cutaneous tumor subtypes: melanoma, \ac{mct}, \ac{scc}, \ac{pnst}, plasmacytoma, trichoblastoma, and histiocytoma. The \acp{wsi} are stored in the pyramidal Aperio file format (\textit{.svs}), allowing direct access to three resolution levels (\SIlist{0.25;1;4}{\micro\meter\per\pixel}).  

In total, the \num{350} \acp{wsi} are accompanied by \annos polygon area annotations. \Cref{tab:annotations} provides an overview of the annotated polygons and the overall annotated area per tissue class. The annotated polygons are provided in the annotation format of the \ac{coco} dataset~\cite{lin2014} as well as an SQLite3 database. For the \ac{coco} format, we have sorted the polygons in increasing order of their hierarchy level, i.e. polygons enclosed by another will be read out after their enclosing polygon. This ordering of polygons can be useful when, for instance, creating annotation masks from the annotation file. These annotation files can also be downloaded from \ac{tcia}~\cite{catch}. 

\begin{table}[H]
\centering
\begin{tabular}{|lS[table-format = 5.0]S[table-format = 5.2]|}
\hline
Class & \multicolumn{1}{l}{Annotation Polygons} & \multicolumn{1}{l|}{Annotation Area [\si{\milli\meter^2}]} \\
\hline
epidermis & 3188 & 2244.57 \\
dermis & 3423 & 16561.82  \\
subcutis & 2850 & 7367.62 \\
bone & 51 & 216.86 \\
cartilage & 16 & 32.15 \\
inflammation \& necrosis & 719 & 2048.54 \\
melanoma & 379 & 6836.61 \\
histiocytoma & 369 & 2941.12 \\
plasmacytoma & 377 & 4750.31 \\
trichoblastoma & 423 & 9072.10 \\
\acl{mct} & 161 & 9329.85 \\
\acl{scc} & 337 & 3513.56 \\
\acl{pnst} & 131 & 11108.78 \\
\hline
total & 12424 & 76023.89 \\
\hline
\end{tabular}
\caption{\label{tab:annotations}Annotated polygons and area per tissue class.}
\end{table}

\subsection*{Dataset visualization}
\begin{figure}[!ht]
\centering
\includegraphics[width=\textwidth]{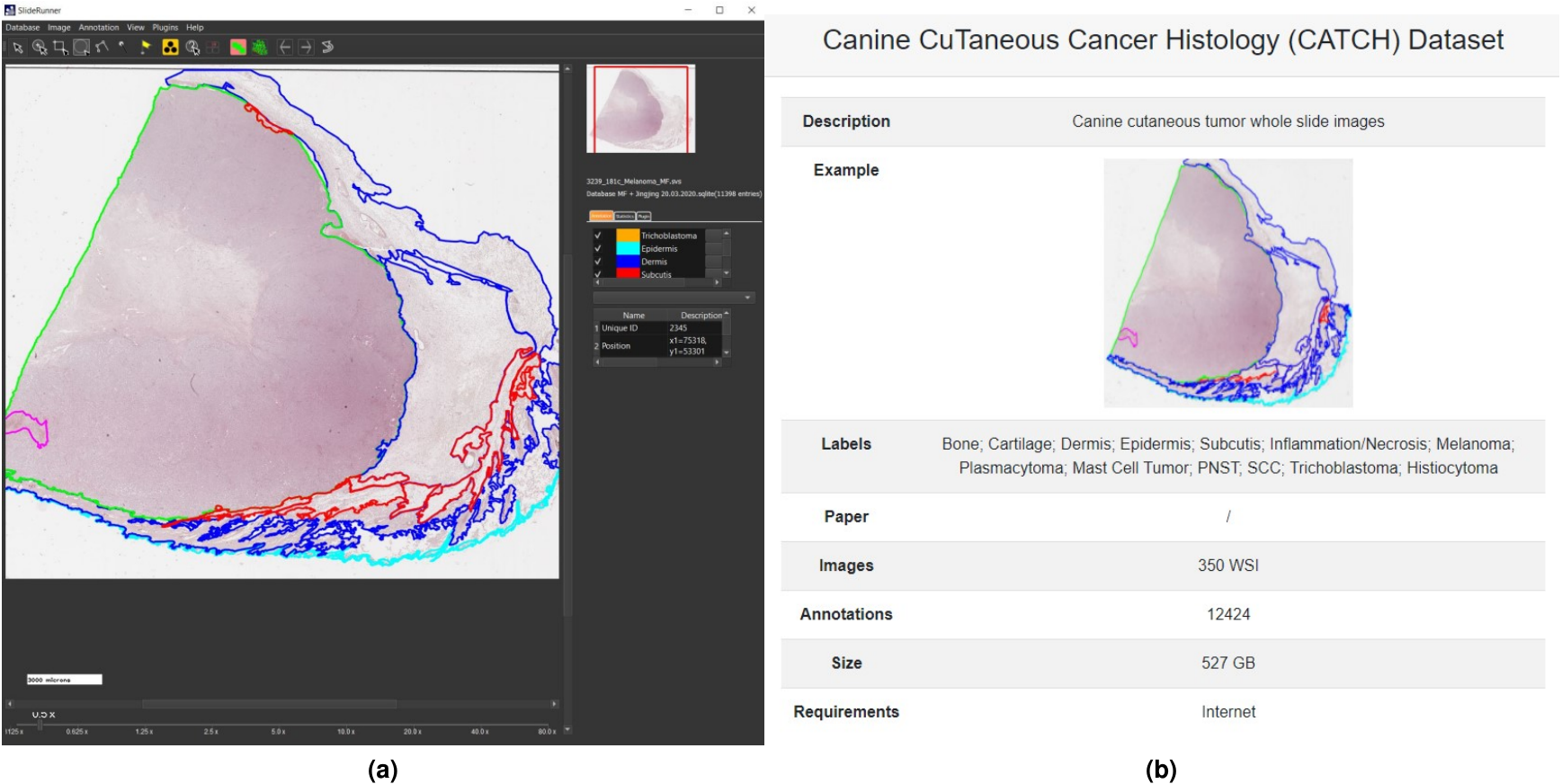}
\caption{User interfaces of recommended open source software tools for dataset visualization. (a) SlideRunner. (b) EXACT.\label{fig:visualization}}
\end{figure}

For visualization of annotations as overlays on top of the original \acp{wsi}, we encourage researchers to use one of the following two alternatives:  

\paragraph{SlideRunner} SlideRunner can be used to visualize the annotations collected in the SQLite3 database. Furthermore, the software allows to set up an additional annotator and extend the database with custom classes and polygon annotations. In our GitHub repository, we provide two code examples to convert SlideRunner annotations into the \ac{coco} format and vice versa. \crefSubFigRef{fig:visualization}{a} illustrates an exemplary \ac{wsi} with pathologist annotations in the SlideRunner user interface.   

\paragraph{EXACT} EXACT enables the collaborative analysis of the dataset with integrated annotation versioning. Furthermore, the REST-API of EXACT allows offline usage and direct interaction with custom machine learning frameworks. The presented dataset can be integrated as a demo dataset into EXACT which enables a direct download of the polygon annotations. Further details can be found in the documentation of EXACT. To make use of additional annotations made by the user, our GitHub repository provides a code example to convert EXACT annotations into the \ac{coco} format. \crefSubFigRef{fig:visualization}{b} shows an overview of the demo dataset in EXACT.

\section*{Technical Validation}
\subsection*{Validation of annotations}
\begin{figure}[!ht]
\centering
\includegraphics[width=\textwidth]{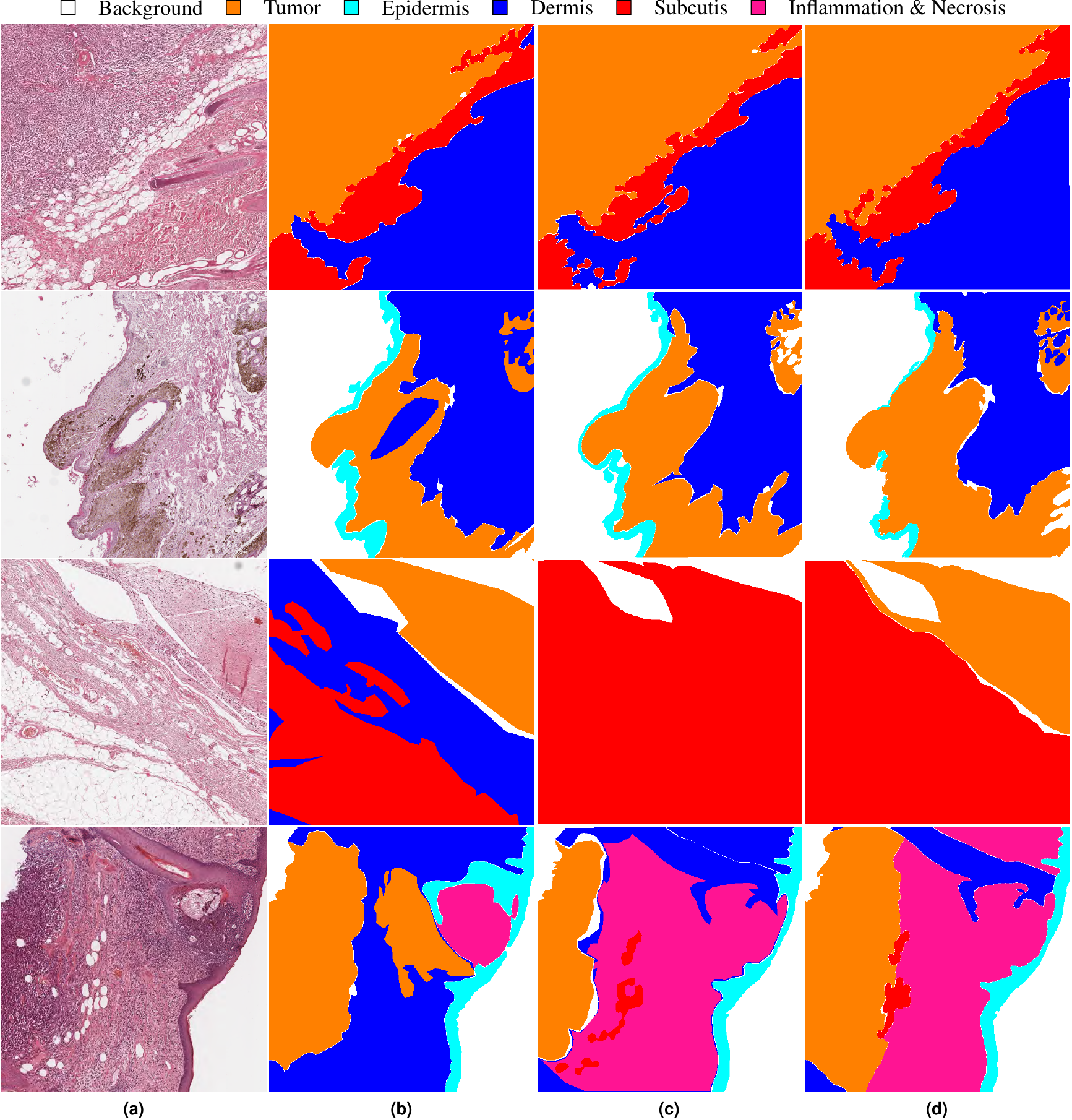}
\caption{Inter-rater variability for four exemplary test patches. (a) Original patch. (b)-(d) Annotations of pathologists. The first two rows show examples with a high inter-rater concordance and the second two rows examples with a low inter-rater concordance. The third example shows different definitions for dermis (blue) and subcutis (red) whilst the fourth example shows a high variation for inflammation \& necrosis (pink). \label{fig:inter-annotator-variability}}
\end{figure}

After database collection, we ensured database consistency by using EXACT to check for and remove annotation duplicates, which occurred in rare cases due to different annotation versions. Previous work on inter-rater variability for contour delineation has demonstrated multiple influence factors on annotator disagreement for this task, such as the complexity of the medical pathology itself but also the hand-eye coordination skills of the raters~\cite{joskowicz2019}. Furthermore, a high level of inter-observer variability can significantly impact the performance of deep learning-based algorithms~\cite{wilm2021}. Therefore, we evaluated the inter-observer variability for the presented dataset with the help of annotations by two additional veterinary pathologists. Even though the comparison of three annotators might only provide an estimate of the full range of inter-observer variability~\cite{joskowicz2019}, it shows the strengths and weaknesses of the provided dataset and highlights annotation classes where computer-aided systems might be of great use to pathologists. Due to the extensiveness of our dataset, we have limited the additional annotations to a \SI{2048}{\micro\meter} $\times$ \SI{2048}{\micro\meter}-sized \ac{roi} on each of the 70 test \acp{wsi}. The size of \SI{2048}{\micro\meter} $\times$ \SI{2048}{\micro\meter} corresponds to the patch size used for training the segmentation algorithm elaborated in the subsequent section. For the selection of these \acp{roi}, we used a uniform sampling across annotation classes to counteract the class imbalance within our dataset. We then positioned the \acp{roi} on a randomly selected vertex of a polygon of the chosen class to explicitly choose tissue boundaries where inter-annotator variability becomes most apparent. \Cref{fig:inter-annotator-variability} visualizes four exemplary patches with a high inter-rater agreement for the first two examples and a low inter-rater agreement for the second two. For quantitative evaluation of the inter-annotator variability, we computed $CI_{pair}$~\cite{kouwenhoven2009} as the average pair-wise Jaccard similarity coefficient for each unique pair of raters $i,\ j$ and the generalized conformity index $CI_{gen}$~\cite{kouwenhoven2009} defined as:

\begin{equation}
CI_{pair,\ c} =  \frac{2}{C(C-1)} \sum_{pairs \hspace*{1pt} i j} \hspace*{3pt} \frac{ A_{c,i} \hspace*{3pt} \cap \hspace*{3pt} A_{c,j}}{A_{c,i} \hspace*{3pt} \cup \hspace*{3pt} A_{c,j}} 
\hspace{1cm}
CI_{gen,\ c} =  \frac{\sum_{pairs \hspace*{1pt} i j} \hspace*{3pt} A_{c,i} \hspace*{3pt} \cap \hspace*{3pt} A_{c,j}}{\sum_{pairs \hspace*{1pt} i j}\hspace*{3pt} A_{c,i} \hspace*{3pt} \cup \hspace*{3pt} A_{c,j}},
\end{equation}%
where $A_c$ are the pixels annotated as class $c \in C$. These two measures have similar values with the same mutual variability between raters, but highly differ when the delineations of one rater are considerably different from the other raters~\cite{kouwenhoven2009}. \Cref{tab:gci} summarizes the pair-wise Jaccard coefficients for each unique pair of raters together with $CI_{pair}$ and $CI_{gen}$ for all annotated tissue classes. The small differences between $CI_{pair}$ and $CI_{gen}$ show that deviations of rater 1, who provided the annotations for the complete dataset, fall within the mutual variability of all raters.

Tumor delineation is a routine task for all pathologists and their extensive experience in this task might be the reason for the comparably high agreement on the tumor annotations with a $CI_{gen,\ tumor}$ of \num{0.8514}. The epidermis is the uppermost layer of the skin and therefore always located at the tissue rim. Furthermore, it is visually distinctly demarcated from the subsequent dermis tissue. These unique characteristics of the epidermis ease the annotation task and might be responsible for the comparably high inter-observer concordance indicated by a $CI_{gen,\ epidermis}$ of \num{0.7512}. The annotators showed a higher inter-rater variability for the two subsequent layers of the healthy skin -- the dermis and subcutis. A closer evaluation of the class-wise confusions showed that these lower scores mostly resulted from mix-ups between these two classes. Such an example is also illustrated in the third example in \cref{fig:inter-annotator-variability}. When combining these two annotation classes into one, the generalized conformity index increased from \num{0.7169} for dermis and \num{0.5836} for subcutis to \num{0.8176} for the combined class. Whereas tumor segmentation is of high relevance for most diagnostic purposes and therefore requires precise definition criteria, we do not see the same relevance for the separation of dermis and subcutis. Thus, we argue that a high inter-rater variability for these tissue classes does not lower the diagnostic interpretability of a segmentation algorithm trained with annotations biased by how these two classes were defined. 

With a generalized conformity index of \num{0.3302}, the concordance for inflammation and necrosis was particularly low. These results, however, were not surprising as these structures are typically far less distinctly demarcated from surrounding tumor tissue due to two biological concepts that have to be considered: Firstly, necrotic areas can frequently be found within tumors where angiogenesis could not keep up with the aggressive growth of the tumor. Furthermore, secondary inflammations can be observed within tumors or at the tumor margin due to the immune system reacting to the neoplasm. Both of these biological mechanisms can result in areas that exhibit neoplastic as well as necrotic or inflammatory characteristics, which makes a precise separation from tumor tissue difficult. Such an example is shown in the last row of \cref{fig:inter-annotator-variability}. Here, all pathologists have annotated an inflamed region located next to the outermost epidermis. Whereas pathologist 1 has annotated the adjacent region as tumor, pathologists 2 and 3 have extended the inflamed region to the rim of the tumor region in the left part of the patch. This tumor region has been delineated similarly by pathologists 1 and 3, whereas pathologist 2 has annotated this region much slimmer. The comparably low conformity index for this class shows the difficulty of clearly separating tissue areas that show a transition between two classes. This limitation of the provided annotations should be considered when evaluating the segmentation results of algorithms trained on the presented dataset.

Overall, the experiments show a high inter-observer agreement for tumor vs. non-tumor, which is highly relevant for most tasks in histopathology. However, they also highlight the difficulty of accurately demarcating necrotic or inflammatory reactions from the surrounding tumor cells. At the same time, it has to be considered that the selected regions for the inter-observer experiments were deliberately placed at tissue transitions and thereby rather over- than underestimate the inter-observer variability. Taking into account that the provided annotations mainly consisted of large connected tissue areas with little tissue interaction, we expect a considerably higher agreement for the complete dataset.

\begin{table}[H]
\centering
\begin{tabular}{|lS[table-format = 1.4]S[table-format = 1.4]S[table-format = 1.4]S[table-format = 1.4]S[table-format = 1.4]|}
\hline
Class & \multicolumn{1}{l}{$CI_{r1, r2}$} & \multicolumn{1}{l}{$CI_{r1, r3}$} & \multicolumn{1}{l}{$CI_{r2, r3}$}&
\multicolumn{1}{l}{$CI_{pairs}$}&
\multicolumn{1}{l|}{$CI_{gen}$} \\
\hline
background & 0.9130 & 0.9239 & 0.9319 & 0.9229 & 0.9229 \\
tumor & 0.8501 & 0.8506 & 0.8535 & 0.8514 & 0.8514 \\
epidermis & 0.8025 & 0.7228 & 0.7270 & 0.7508 & 0.7512 \\
dermis & 0.6875 & 0.7015 & 0.7659 & 0.7183 & 0.7169 \\
subcutis & 0.5254 & 0.5872 & 0.6421 & 0.5849 & 0.5836 \\
inflammation \& necrosis & 0.2823 & 0.3040 & 0.3895 & 0.3253 & 0.3302 \\
\hline
\end{tabular}
\caption{\label{tab:gci}Class-wise conformity index computed for all unique pairs of annotators. $CI_{pairs}$ averages the pair-wise conformity indices whereas $CI_{gen}$ is a generalized version of the Jaccard coefficient.}
\end{table}

\subsection*{Dataset validation through algorithm development}
\begin{figure}[!ht]
\centering
\includegraphics[width=\textwidth]{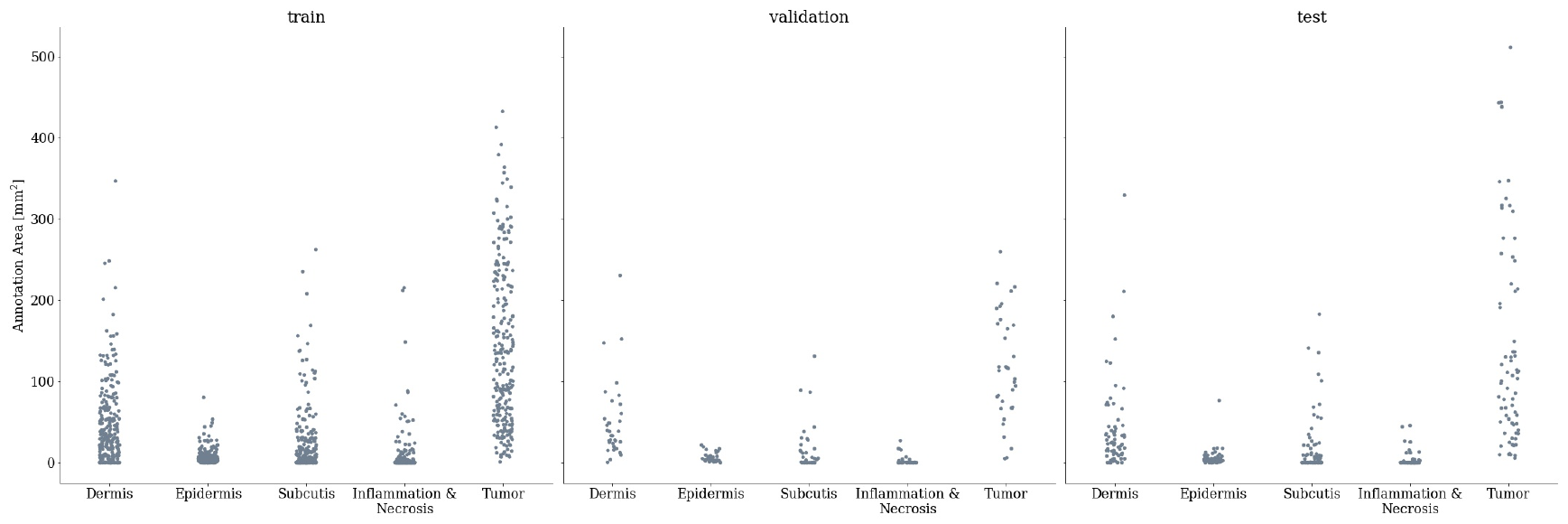}
\caption{Distribution of annotation area per class across dataset splits. The train split consisted of 245 \aclp{wsi}, the validation set of 35 \aclp{wsi} and the test set of 70 \aclp{wsi}.\label{fig:dataset_split}}
\end{figure}

For further validation of the dataset, we evaluated two \ac{cnn} architectures for the task of tissue segmentation and tumor subtype classification. For both tasks, we used the same dataset split: For each of the seven tumor subtypes, we randomly selected 35 \acp{wsi} for training, five for validation, and ten for testing. Thereby, we ensured equal distribution of tumor subtypes in each split. Even though \acp{wsi} from the same canine patient showed different tissue sections, we maintained a dataset split at patient level to avoid data leakage. \Cref{fig:dataset_split} visualizes the distribution of annotated area per class across the \acp{wsi} of the train, validation, and test split. For simplicity, we have combined all tumor subtypes into one class for tumor segmentation, and consider the tumor subtypes separately only during tumor classification. The visualization shows similar distributions for all splits, which ensures that our test set evaluations are representative for the complete data distribution. However, the distributions also highlight the high class imbalance within the dataset which has to be considered during the development of algorithms for computer-aided tasks. A detailed overview of the slide-level split can be obtained from the GitHub repository in form of a \ac{csv} table together with code for implementing the \ac{cnn} architectures presented in the subsequent sections.

\subsubsection*{Tissue segmentation}
\begin{figure}[!ht]
\centering
\includegraphics[width=0.5\textwidth]{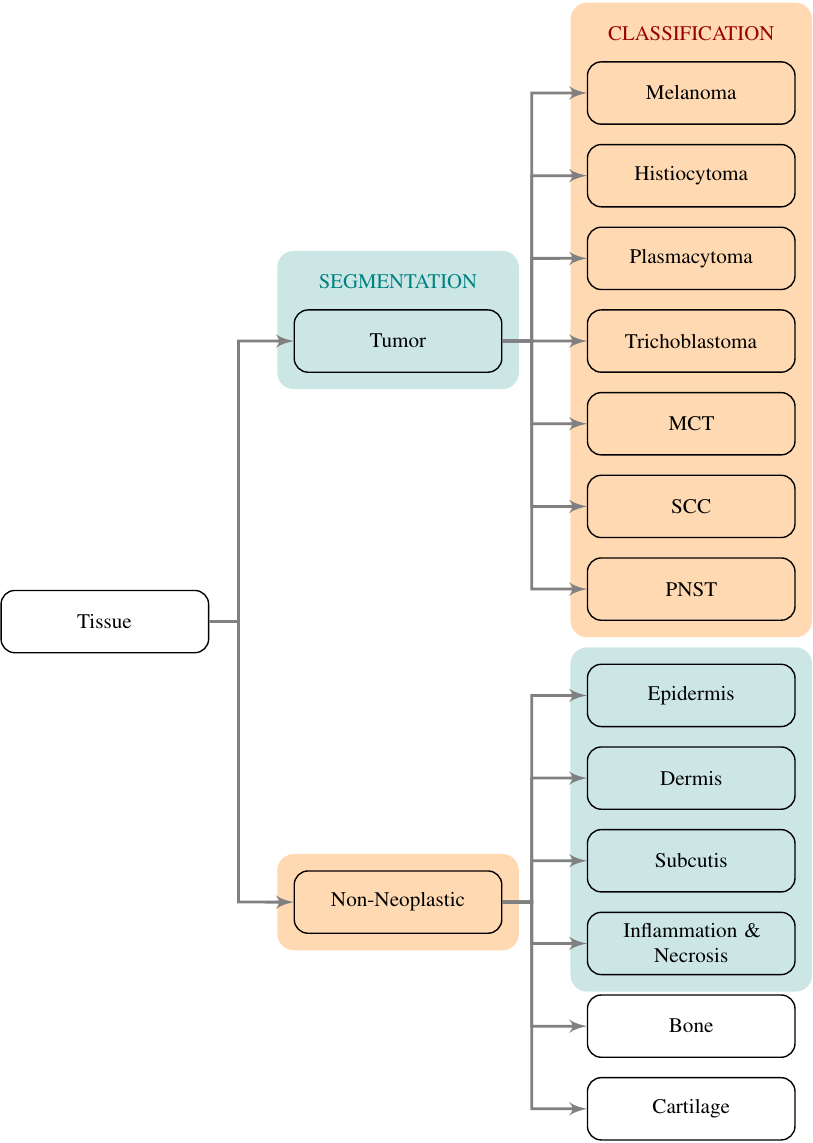}
\caption{Taxonomy of tissue classes. Classes highlighted in green were used for training the segmentation network and classes highlighted in orange were used to train the tumor subtype classification network. \acs{mct}: \acl{mct}, \acs{scc}: \acl{scc}, \acs{pnst}: \acl{pnst}.\label{fig:tissue_classes}}
\end{figure}

For the task of tissue segmentation, we trained a UNet~\cite{ronneberger2015} to distinguish between four non-neoplastic tissue classes (epidermis, dermis, subcutis, and inflammation combined with necrosis) and all tumor subtypes combined into one \say{tumor} label. These five classes were accompanied by a sixth \say{background} class. For this background class, we used Otsu's adaptive thresholding~\cite{otsu1979} to compute a \say{white} threshold for each slide and assigned the \say{background} label to all non-annotated pixels that exceeded this \say{white} value. Overall, this resulted in six classes used for training the segmentation network. \Cref{fig:tissue_classes} visualizes the annotation taxonomy and highlights the classes used for segmentation in green. Due to the low diagnostic significance and limited availability of bone and cartilage annotations, we excluded these classes from training and evaluating the proposed methods. Non-annotated tissue areas were also excluded from training and evaluation.  

For segmentation, we used the fastai~\cite{howard2020} UNet implementation with a ResNet18~\cite{he2016} backbone pre-trained on ImageNet~\cite{russakovsky2015}. Image patches of \num{512} $\times$ \num{512} \si{pixels} at a resolution of \SI{4}{\micro\meter\per\pixel} (2.5X), which corresponds to a tissue size of \num{2048} $\times$ \num{2048} \si{\micro\meter^2}, were used as input. We decided to use this 16-fold down-sampled resolution because input patches then covered more context, which has shown to be more beneficial for segmentation results in previous work~\cite{sirinukunwattana2018} and was confirmed by initial experiments on the validation dataset. To limit the number of non-informative white background patches and overcome class imbalances with random sampling, we propose an adaptive patch-sampling strategy: For each slide, we initialized the class probabilities as a uniform distribution over all annotation classes used on the respective slide. For a fixed number of training patches, we first sampled a class according to the class probabilities and then randomly selected a position within one of the polygons of this class. The final training patch was centered at this pixel location. We refer to this guided selection of a fixed number of patches as \say{pseudo-epoch}~\cite{bertram2019}. After each pseudo-epoch, the model performance was evaluated on a fixed number of validation patches sampled in a similar fashion. The model performance was assessed using the class-wise Jaccard similarity coefficient $J_c$. Prior to the next pseudo-epoch, we updated the class-wise probabilities $p_c$ of each slide according to the complement of the corresponding class-wise Jaccard coefficient $J_c$:

\begin{equation}
\label{eq:class-probability}
    p_c = 1 - J_c = 1- \left(\frac{1}{N_v}\sum_{i=1}^{N_v} J_{c,i}\right)
    \hspace{1cm}
    N_v: \text{number of validation patches}.
\end{equation}%
This adaptive sampling strategy aimed for a faster convergence by over-sampling classes where the model faced difficulties. For each pseudo-epoch, we sampled ten patches per slide, resulting in \num{2450} training patches and \num{350} validation patches per pseudo-epoch. All patches were normalized using the RGB statistics of the training set, i.e. subtract the mean and divide by the standard deviation of all tissue-containing areas of the training \acp{wsi}. We trained the model for up to 100 pseudo-epochs and selected the configuration with the highest class-averaged Jaccard coefficient on the validation patches. We trained with a maximal learning rate of \num{e-4}, a batch size of four, and used discriminative fine-tuning~\cite{howard2018} provided by the fastai package. During training, online data augmentation was used, composed of random flipping, affine transformations, and random lightning and contrast change. To meet the class imbalance within the data, the model was trained with a combination of the generalized Dice loss~\cite{sudre2017} and the categorical focal loss~\cite{lin2017}.

\begin{figure}[!ht]
\centering
\includegraphics[width=\textwidth]{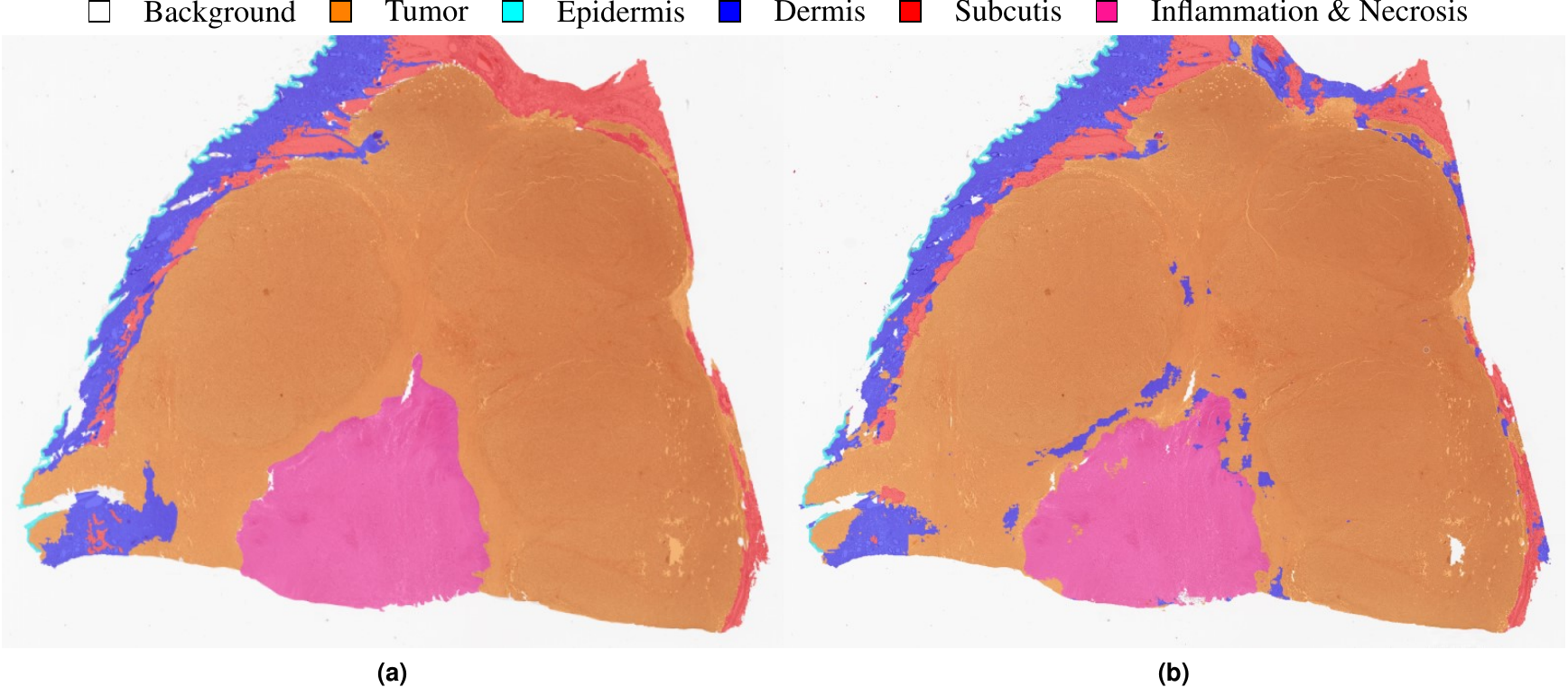}
\caption{Exemplary segmentation result. (a) Annotation. (b) Prediction.\label{fig:segmentation}}
\end{figure}

\begin{figure}[!ht]
\centering
\includegraphics[width=0.7\textwidth]{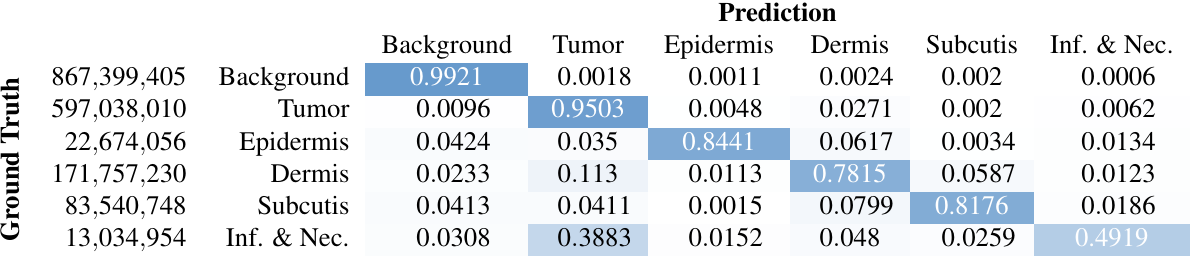}
\caption{Segmentation confusion matrix (pixel-based). The numbers on the left summarize the pixel-count per class.\label{fig:cm_segmentation}}
\end{figure}

After model training, we computed a slide segmentation output using a moving-window patch-wise inference with an overlap of half the patch size, i.e. \num{256} pixels. In the overlap area, we averaged the class probabilities computed as softmax-output of the model predictions. This inference resulted in a three-dimensional output tensor with the slide dimensions in x- and y-direction and the number of segmentation classes in z-direction. The per-pixel labels were then computed as the class with the maximum entry in z-direction. \cref{fig:segmentation} visualizes an exemplary segmentation result with the original slide and annotated regions on the left and the predicted segmentation output on the right. For quantitative performance evaluation, we accumulated the pixel-based confusion matrices of all \acp{wsi} of the test set and then computed the class-wise Jaccard similarity coefficient. \Cref{fig:cm_segmentation} visualizes the row-normalized accumulated confusion matrix for a resolution of \SI{4}{\micro\meter\per\pixel}. The color-coding visualizes the row-wise normalization. The first column of \cref{tab:jaccard} summarizes the class-wise Jaccard coefficients computed from the confusion matrix. Overall, the network scored a class-averaged Jaccard coefficient of \num{0.7047}. Due to the high class imbalance, we also computed a frequency-weighted Jaccard coefficient by multiplying the class-wise coefficients with the class-wise ratio of the respective pixels in the ground truth and summing up over all values. This yielded a frequency-weighted coefficient of \num{0.9001}. The results show that especially for the background and tumor class, the network scored high Jaccard coefficients of \num{0.9757} and \num{0.9044} respectively. This could mainly be attributed to a high sensitivity, i.e. few areas were overlooked. However, the algorithm misclassified a relatively large amount of non-neoplastic pixels as cancerous, especially inflamed and necrotic regions, yielding a comparably low Jaccard coefficient of \num{0.3023} for this combined class. Yet, this behavior meets clinical demands, as the costs of falsely classifying healthy tissue as tumor are far lower than overlooking neoplastic regions which could at worst lead to a false diagnosis. The high amount of neoplastic and inflammatory regions misclassified as tumor can again be ascribed to the necrotic and inflamed regions often interspersed between tumor cells, which makes a clear distinction difficult.  The results of our inter-observer experiments have shown that a precise definition of these classes can be difficult even for trained pathologists. Therefore, algorithmic confusions between these classes should always be evaluated with the above-mentioned challenge in mind.

To evaluate whether training the algorithm on annotations of a single rater introduced a bias towards this rater, we additionally computed \ac{roi} Jaccard coefficients for the test patches included in the inter-rater experiments. These are summarized in \cref{tab:jaccard}. Overall, the results do not show a clear bias towards rater 1 for most classes, as the results fall within the range of the inter-annotator conformity indices. For the combined class of inflammation and necrosis, the algorithm shows a tendency towards rater 1 but still shows a very poor agreement with a Jaccard score of \num{0.2816}. This again highlights the difficulty of accurately defining this class. When comparing the \ac{roi} Jaccard coefficients of rater 1 to the \ac{wsi} Jaccard coefficients, the algorithm shows mostly lower performance, which underlines the increased complexity of the \acp{roi}, which were deliberately placed at tissue transitions.

\begin{table}[H]
\centering
\begin{tabular}{|lS[table-format = 1.4]S[table-format = 1.4]S[table-format = 1.4]S[table-format = 1.4]|}
\hline
Class & \multicolumn{1}{l}{\ac{wsi}} & \multicolumn{1}{l}{\ac{roi}$_{r1}$} & \multicolumn{1}{l}{\ac{roi}$_{r2}$}  & \multicolumn{1}{l|}{\ac{roi}$_{r3}$ }  \\
\hline
background & 0.9757 & 0.9362 & 0.9355 & 0.9430 \\
tumor & 0.9044 & 0.8399 & 0.7911 & 0.8193 \\
epidermis & 0.6661 & 0.7033 & 0.6685 & 0.6167 \\
dermis & 0.6753 & 0.6988 & 0.6285 & 0.6740 \\
subcutis & 0.7043 & 0.6232 & 0.5101 & 0.6435 \\
inflammation \& necrosis & 0.3023 & 0.2816 & 0.1331 & 0.1646 \\
\hline
\end{tabular}
\caption{\label{tab:jaccard}Class-wise Jaccard similarity score for all \acp{wsi} of the test set annotated by rater 1 and the \aclp{roi} evaluated in the inter-rater experiments annotated by raters 1, 2, and 3.}
\end{table}

\noindent To evaluate whether the morphology of certain tumor subtypes within the dataset made a precise differentiation of the tissue classes more difficult, we also computed the class-wise Jaccard coefficients per tumor subtype. These results are summarized in \cref{tab:jaccard_sub_type}. The results show that the network performed exceptionally well for trichoblastoma with a Jaccard coefficient of \num{0.9650} but was challenged by \ac{scc} samples with a Jaccard coefficient of \num{0.7185}. The \ac{scc} confusion matrix revealed that \SI{60.62}{\percent} of the pixels annotated as inflammation or necrosis were falsely classified as tumor. \Ac{scc}, however, is known to cause severe inflammatory reactions~\cite{gasparoto2012}. Infiltration of these inflammatory cells in-between the nests or trabecular of neoplastic epidermal cells can make an accurate distinction of both classes difficult, which could also be seen when evaluating the inter-annotator variability on the presented dataset. 

\begin{table}[H]
\centering
\begin{tabular}{|lS[table-format = 1.4]|}
\hline
Class & \multicolumn{1}{l|}{Tumor \ac{wsi} Jaccard Score} \\
\hline
melanoma & 0.8394\\
histiocytoma & 0.8714\\
plasmacytoma & 0.9419\\
trichoblastoma & 0.9650\\
\acl{mct} & 0.8721\\
\acl{scc} & 0.7185\\
\acl{pnst} & 0.9166\\
\hline
\end{tabular}
\caption{\label{tab:jaccard_sub_type}Tumor Jaccard similarity score computed from the confusion matrix accumulated over all ten test \acp{wsi} of the respective subtype.}
\end{table}

\noindent Recent work has shown that deep learning-based models face difficulties when being applied to \acp{wsi} digitized by a slide scanning system different from the one used for training the algorithm~\cite{moyes2018, aubreville2021}. Due to practical feasibility, a subset of the presented dataset was digitized with a different slide-scanning system. We compensated for this by ensuring a similar distribution of scanner domains in our training and test split and observed similar performances on our test data with mean Jaccard coefficients of \num{0.7026} for the ScanScope CS2 and \num{0.6986} for the AT2. Nevertheless, we are currently creating a multi-scanner dataset of a subset of the data presented in this work and future work will evaluate the transferability of trained models to unseen scanner domains and the development of domain-invariant algorithms.  

\subsubsection*{Tumor classification}
Besides tissue segmentation, we trained an additional \ac{cnn} for tumor subtype classification. For this, an EfficientNet-B0~\cite{tan2019} was trained to distinguish between all seven tumor classes. We combined all non-neoplastic tissues used for training the segmentation network into one rejection class, allowing for inference on patches where no tumor was present. This resulted in eight classes used for training the classification network. Due to the high morphological resemblance of round cell tumors, where cell-level information might be required to distinguish the individual subtypes, we used the original scanning resolution of \SI{0.25}{\micro\meter\per\pixel} (40X) for classification. This corresponds to the diagnostic workflow of pathologists, who would first use a lower resolution to locate the tumor region and then use a higher resolution to classify the tumor. To retain as much context as possible, we increased the patch size to \num{1024} $\times$ \num{1024} \si{pixels}. We used the same train-test split as used for tissue segmentation and trained the network for \num{100} pseudo-epochs with ten patches per slide in each epoch. Fixing this number of sampled patches per slide ensured that each tumor was represented equally and network training was not affected by the very differently sized tumors highlighted by \cref{tab:annotations}, where \ac{pnst} annotations make up for almost \SI{15}{\percent} of the overall annotated area whereas histiocytoma annotations only for about \SI{4}{\percent}. For each slide, we set the probability of sampling a tumor patch seven times higher than the probability of sampling a non-neoplastic patch, as these were present in all slides of the seven tumor types, whereas tumor-specific patches were only available for the training slides of the respective tumor subtype. For the non-neoplastic patches, we ensured an equal sampling of the different tissue classes by first randomly sampling a class and then selecting a patch within one of the polygons of this class. We followed an area-based polygon sampling strategy to ensure an equal distribution of sampled patches across the annotated polygons of the respective class. Furthermore, a patch was only used for training the classification network if at least \SI{90}{\percent} of the pixels were annotated as the sampled class. All patches were normalized using the training set statistics. Similar to the segmentation network, we used online data augmentation. The network was trained with a batch size of four and a maximal learning rate of \num{e-3}. We used the Adam optimizer and trained the model with cross-entropy loss. We used the mean patch-level accuracy to guide the model selection process.

\begin{figure}[!ht]
\centering
\includegraphics[width=\textwidth]{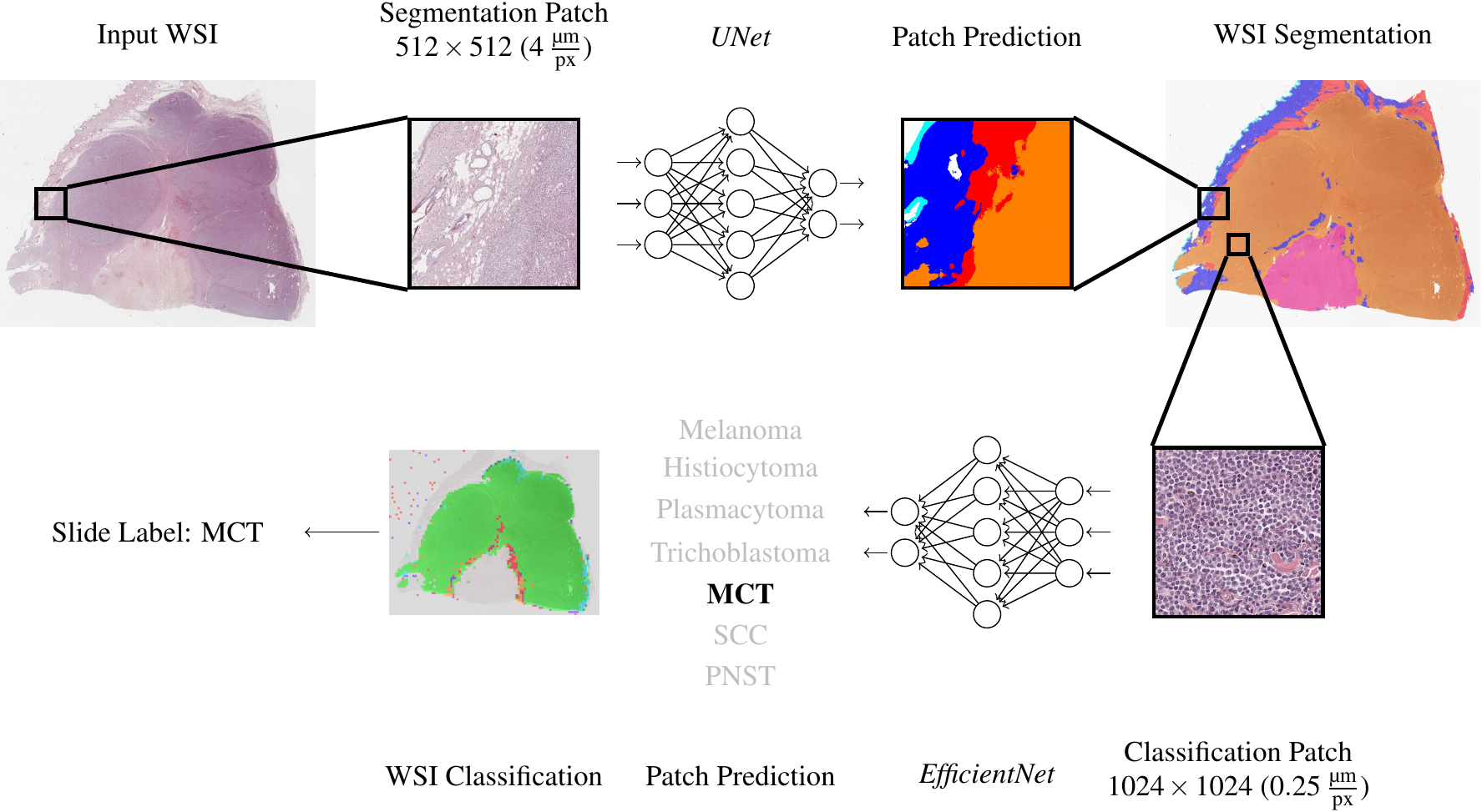}
\caption{Patch segmentation and classification pipeline. Due to different resolutions and patch sizes between the segmentation and the classification task, a single segmentation patch holds multiple classification patches. Only patches segmented as tumor are classified into a tumor subtype. \acs{mct}: \acl{mct}, \acs{scc}: \acl{scc}, \acs{pnst}: \acl{pnst}.\label{fig:pipeline}}
\end{figure}

To combine the pixel-wise segmentation with the patch-wise tumor subtype classification, we propose the following slide inference pipeline, visualized in \cref{fig:pipeline}: First, a \ac{wsi} is segmented into six tissue classes, using the segmentation network described in the previous subsection. The spatial resolution of the pixel-wise segmentation map corresponds to the \ac{wsi} at the chosen resolution of \SI{4}{\micro\meter\per\pixel}, which represents a \num{16}-fold down-sampling in each dimension. This segmentation map is up-sampled to the original resolution and only patches that were fully segmented as tumor obtain a patch label by the tumor subtype classification network. These patch classifications are then combined into a slide label by using majority voting. By training the tumor subtype classification network on an additional rejection class comprised of non-neoplastic tissue, we aimed to compensate for false-positive tumor segmentation predictions. If the classification network assigned the rejection label for these patches, they were excluded from the subsequent majority voting. Inference time for this pipeline was measured using an NVIDIA Quadro RTX 8000 graphics processing unit. \Ac{wsi} segmentation took \SI{15\pm 7}{\sec} (\si{\mu \pm \sigma}) for an average of \SI{405\pm 177}{patches} ($\hat{=}$ \SI{37}{\milli\sec} per patch). In our two-stage inference pipeline, only patches from areas segmented as tumor were passed on to the tumor subtype classification network. This significantly reduced the number of patches to be predicted, however, due to the higher resolution of the classification network, we still measured inference times of \SI{6\pm 5}{\minute} for classification with an average of \SI{2472\pm 1873}{patches} per slide ($\hat{=}$ \SI{155}{\milli\sec} per patch). The comparatively high variance within these inference times resulted from the high variance of tissue and tumor area within the test set. On average, the \acp{wsi} were sized \SI{6.47 \pm 2.89 e+9}{\pixel} at the original resolution.

\begin{figure}[!ht]
\centering
\includegraphics[width=\textwidth]{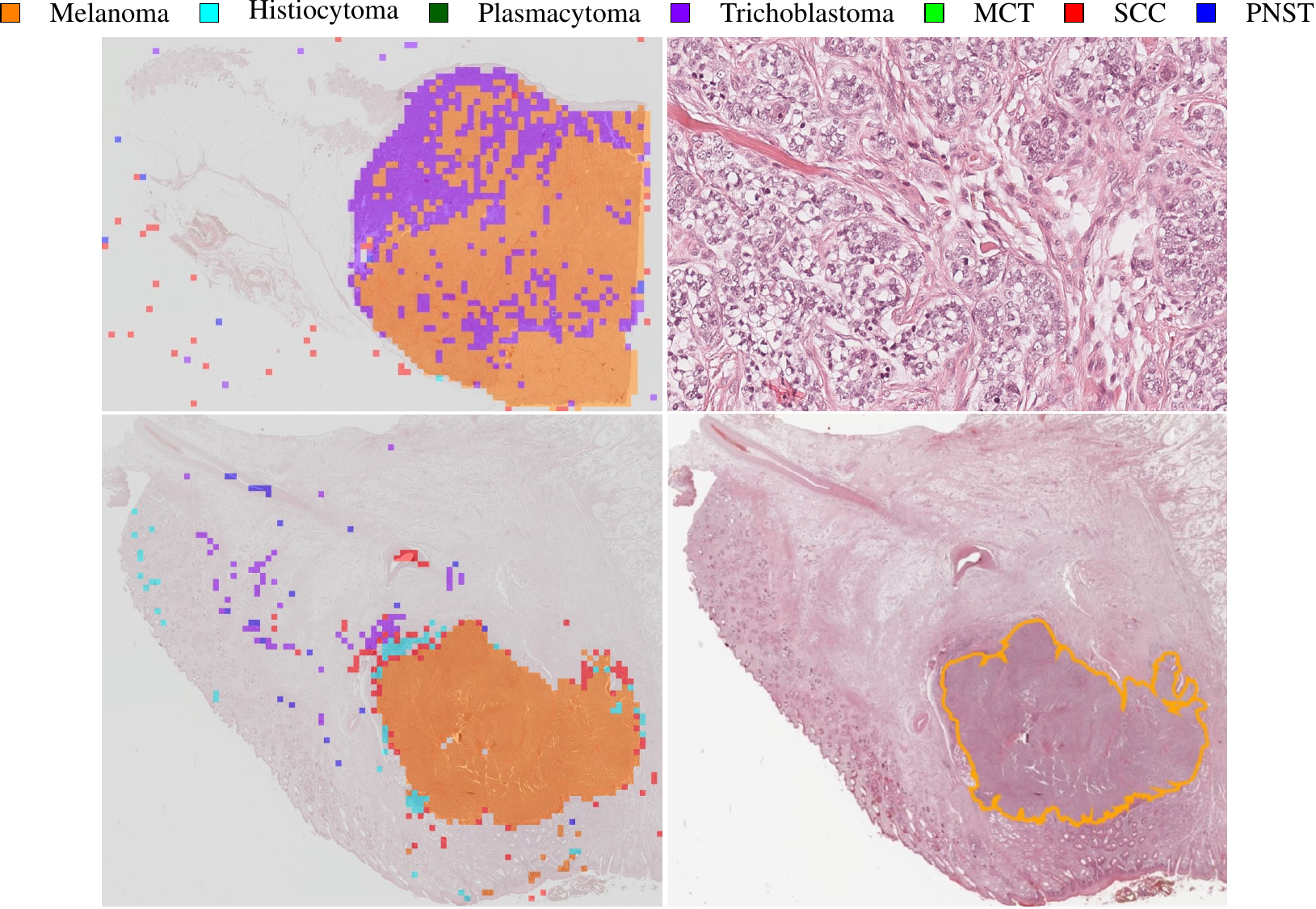}
\caption{Exemplary classification results. The upper example shows a trichoblastoma sample misclassified as melanoma with the classification output on the left and the magnified tumor region on the right. The tumor region shows a high number of pleomorphic tumor cells. The lower example shows a melanoma sample with the classification output on the left and the annotated sample on the right. The classification output shows a high ratio of misclassified patches caused by a false tumor prediction during segmentation. \acs{mct}: \acl{mct}, \acs{scc}: \acl{scc}, \acs{pnst}: \acl{pnst}.\label{fig:classification}}
\end{figure}

When applying the slide inference pipeline to all 70 \acp{wsi} from the test set, we classified 69 \acp{wsi} correctly, yielding a slide classification accuracy of \SI{98.57}{\percent}. The misclassified slide is depicted in the upper example of \cref{fig:classification}. Here, the model falsely labeled a trichoblastoma slide as melanoma. A closer examination of this slide revealed a high number of undifferentiated, pleomorphic cells, i.e. tumor cells of varying shapes and sizes, visualized in the magnified tumor region on the upper right side of \cref{fig:classification}. The region shows characteristics of epithelial tumors, the superordinate tumor category of trichoblastomas, but melanomas, too, can be composed of epitheloid cells. Melanomas are typically highly pleomorphic, which might have caused the misclassification as melanoma. The upper example in \cref{fig:classification} also shows that some misclassified patches are located on the white \ac{wsi} background. A closer look at these areas revealed small parts of detached tissue or dust artifacts, which were mistaken as tumor by the segmentation network and then falsely passed on to the classification network. This could be circumvented by additionally training the classification network on background patches or applying a post-processing step such as morphological closing to the segmentation output.

\begin{figure}[!ht]
\centering
\includegraphics[width=0.9\textwidth]{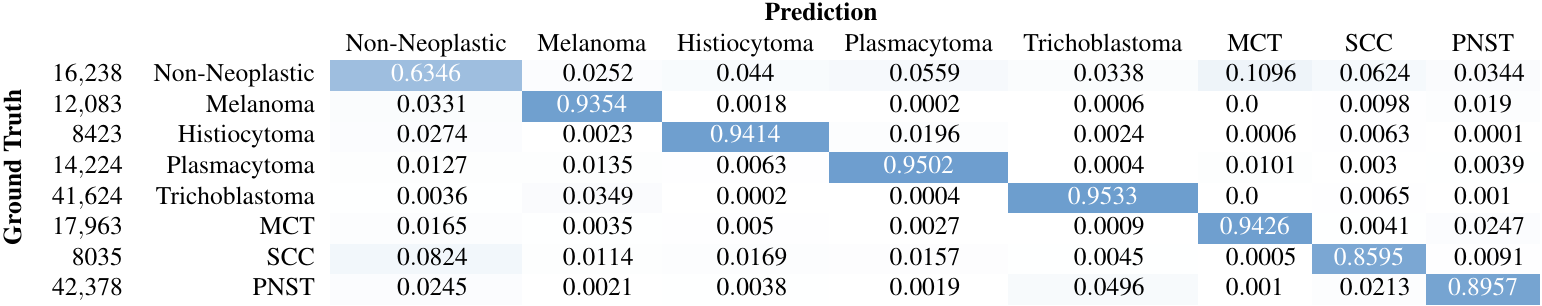}
\caption{Classification confusion matrix (patch-based). The numbers on the left summarize the patch-count per tumor class.\label{fig:cm_classification}}
\end{figure}

To evaluate whether some tumor subtypes were more difficult for the classification network than others, i.e. the majority voting was affected by many false patch classifications, we evaluated the confusion matrix of the tumor subtype patch classification, of which a row-normalized version is shown in \cref{fig:cm_classification} with the color-coding again representing the normalization. This confusion matrix only includes patches that were segmented as tumor and thereby passed on to the classification network. The first row of the matrix shows that the segmentation network passed on \num{16238} false-positive tumor patches to the classification network of which \SI{63.46}{\percent} were recovered by the rejection class. From the remaining rows, we computed tumor-wise recalls and precisions, i.e. the ratio of all patches correctly classified as the respective subtype to all patches labeled or respectively predicted as the corresponding subtype. These metrics, summarized in \cref{tab:metrics}, only consider confusion among tumor subtypes and not with the non-neoplastic class. The confusion matrix and the results in \cref{tab:metrics} show that \ac{scc} generally was the most difficult class for the network to distinguish. Looking at the results in detail, however, the comparably low F$_1$~score of 0.8773 can mostly be attributed to the low classification precision, meaning a lot of tumor patches were falsely classified as \ac{scc}. A closer look at the classification outputs showed that these misclassifications were mostly located at the tumor boundaries. This observation could be linked to the severe inflammatory reactions that are typically caused by \acp{scc}~\cite{gasparoto2012}. During training, inflammatory reactions to tumor growth might have been more common for \ac{scc} samples than for other subtypes, which might have caused the model to mistake the interaction of tumor and inflammatory cells as \ac{scc}.

\begin{table}[H]
\centering
\begin{tabular}{|lS[table-format = 1.4]S[table-format = 1.4]S[table-format = 1.4]|}
\hline
Class & \multicolumn{1}{l}{Precision} & \multicolumn{1}{l}{Recall} & \multicolumn{1}{l|}{F$_1$~score}\\
\hline
melanoma & 0.8555 & 0.9675 & 0.9081\\
histiocytoma & 0.9399 & 0.9679 & 0.9537\\
plasmacytoma & 0.9684 & 0.9624 & 0.9654\\
trichoblastoma & 0.9478 & 0.9568 & 0.9522\\
\acl{mct} & 0.9886 & 0.9584 & 0.9732\\
\acl{scc} & 0.8251 & 0.9367 & 0.8773\\
\acl{pnst} & 0.9783 & 0.9182 & 0.9473\\
\hline
\end{tabular}
\caption{\label{tab:metrics}Patch-level tumor precision, recall and F$_1$~score per tumor subtype.}
\end{table}

\noindent The lower example in \cref{fig:classification} shows a melanoma sample where the majority voting yielded the correct classification label but was affected by many false patch classifications. A comparison to the ground truth annotations depicted on the lower right side of \cref{fig:classification} reveals that this can be rooted back to a false tumor prediction of the preceding segmentation network, as only the area correctly classified as melanoma was also annotated as tumor. For this example, the rejection class could not fully recover the errors made by the segmentation network. Even though the majority voting resulted in the correct slide label for this example, one should always take into consideration some measure of confidence for the majority voting, indicating how difficult the final decision was. Determining the patch-level entropy, for instance, could highlight slides where the distribution of patch classifications across the tumor subtypes resembled a uniform distribution, i.e. scored a high entropy and the decision was made less confidently. This entropy could then be used for weighted voting of the slide label instead of a simple majority voting.

\subsubsection*{Dataset insights from algorithm development}
Overall, the algorithm results on the provided database validate database quality as a successful training of a segmentation algorithm on the dataset proves the consistency of the annotations. When comparing the algorithm to the annotations of rater 1, the Jaccard scores fall within the range of inter-annotator concordance, indicating that the provided annotations did not introduce a bias into algorithm development. Furthermore, the experiments highlighted strengths and weaknesses of the provided dataset, as for instance \acp{scc} are more affected by inflammatory reactions, which makes them less suited for training an algorithm for a clear distinction of tumor and inflammation. 

\section*{Usage Notes}
All code examples are based on OpenSlide~\cite{goode2013} for \ac{wsi} processing and fastai~\cite{howard2020} for network training. To apply the fastai modules to \acp{wsi}, we provide custom data loaders in our GitHub repository. The annotation and visualization tools used for this work -- SlideRunner and EXACT -- are both open source and can be downloaded from the respective GitHub repositories.

\section*{Code Availability}
Code examples for training the segmentation and classification architectures can be found in the form of Jupyter notebooks in our GitHub repository (\href{https://github.com/DeepPathology/CanineCutaneousTumors}{https://github.com/DeepPathology/CanineCutaneousTumors}). Furthermore, we provide exported fastai learners to reproduce the results stated in this work. The \textit{datasets.csv} file lists the train, validation, and test split on slide level. For network inference, we provide two Jupyter notebooks for patch-level results (\textit{segmentation\_inference.ipynb} and \textit{classification\_inference.ipynb}) and one notebook for slide-level results. This \textit{slide\_inference.ipynb} notebook produces segmentation and classification outputs as compressed numpy arrays. After inference, these prediction masks can be visualized as overlays on top of the original images using our custom SlideRunner plugins \textit{wsi\_segmentation.py} and \textit{wsi\_classification.py}. To integrate these plugins into their local SlideRunner installation, users have to copy the respective plugin from our GitHub repository into their SlideRunner \textit{plugin} directory. Additionally, the \textit{slide\_inference.ipynb} notebook provides methods to compute confusion matrices from network predictions and calculate class-wise Jaccard coefficients and the tumor classification recall. As mentioned previously, we provide six python modules to convert annotations back and forth between \ac{coco} and EXACT, \ac{coco} and SQLite, and EXACT and SQLite formats. This enables users to extend the annotations by custom classes or polygons in their preferred annotation format. These modules can be found in the \textit{annotation\_conversion} directory of our GitHub repository. 

\bibliography{bibliography}

\section*{Author contributions statement}
F.W. wrote the manuscript, carried out data analysis and developed the presented algorithms. F.W., K.B. and M.A. all contributed to method development. M.F. provided annotations for \SI{82}{\percent} of the data, reviewed the remaining annotations for correctness and completeness and provided medical expertise for discussion of the results. C.M. and J.Q. contributed to data collection and method development. C.P. and L.D. provided manual annotations for selected regions on the test data to evaluate inter-observer variability. C.B. and R.K. provided medical and A.M., K.B. and M.A. technical expertise. All authors contributed to the preparation of the manuscript and approved of the final manuscript for publication.

\section*{Competing interests}
The authors declare no competing interests.

\begin{acronym}
\acro{catch}[CATCH]{CAnine cuTaneous Cancer Histology}
\acro{ihc}[IHC] {immunohistochemical}
\acro{he}[H\&E]{Hematoxylin \& Eosin}
\acro{wsi}[WSI]{whole slide image} 
\acro{bach}[BACH]{Grand Challenge on BreAst Cancer Histology images}
\acro{paip}[PAIP]{Pathology Artificial Intelligence Platform}
\acro{adp}[ADP]{Atlas of Digital Pathology}
\acro{bracs}[BRACS]{BReAst Carcinoma Subtyping}
\acro{droid}[DROID]{Diagnostic Reference Oncology Imaging Database}
\acro{fcn}[FCN]{fully convolutional network}
\acro{tcia}[TCIA]{The Cancer Imaging Archive}
\acro{mct}[MCT]{mast cell tumor}
\acro{pnst}[PNST]{peripheral nerve sheath tumor} 
\acro{scc}[SCC]{squamous cell carcinoma}
\acro{coco}[MS COCO]{Microsoft Common Objects in Context}
\acro{cnn}[CNN]{convolutional neural network}
\acro{csv}[\textit{.csv}]{comma-separated value}
\acro{roi}[ROI]{region of interest}

\end{acronym}

\end{document}